\documentclass[pra,twocolumn,preprintnumbers,amsmath,amssymb,superscriptaddress]{revtex4-2}

\usepackage{color}
\usepackage{graphicx}% Include figure files
\usepackage{epstopdf}
\usepackage{epsfig}
\usepackage{dcolumn}% Align table columns on decimal point
\usepackage{bm}% bold math
\usepackage{hyperref}
\usepackage{mathrsfs}
\usepackage{comment}
\usepackage{soul}
\usepackage[normalem]{ulem} %to strike the words
\usepackage{ulem}
\usepackage{xcolor}
\usepackage{amsfonts}
\usepackage{amsmath}
\usepackage{amsxtra}
\usepackage{amssymb}
\usepackage{mathrsfs}
\usepackage{hyperref}
\usepackage{soul}
\usepackage{float}
\usepackage{relsize}
\usepackage{placeins}
\usepackage[T1]{fontenc}
\begin{document}

%\preprint{APS/123-QED}

\title{Optical and atomic decoherence in entangled atomic ensembles generated by quantum nondemolition measurements}

\author{Shuai Gao}
\thanks{The indicated authors are joint first authors}
\affiliation{Joint International Research Laboratory of Information Display and Visualization, School of Electronic Science and Engineering, Southeast University, Nanjing 210096, China} 
\affiliation{State Key Laboratory of Precision Spectroscopy, School of Physical and Material Sciences, East China Normal University, Shanghai 200062, China}

\author{Shuang Li}
\thanks{The indicated authors are joint first authors}
\affiliation{State Key Laboratory of Precision Spectroscopy, School of Physical and Material Sciences, East China Normal University, Shanghai 200062, China}

\author{Manish Chaudhary}
\affiliation{State Key Laboratory of Precision Spectroscopy, School of Physical and Material Sciences, East China Normal University, Shanghai 200062, China}
\affiliation{New York University Shanghai, 1555 Century Ave, Pudong, Shanghai 200122, China}

\author{Matthew Prest}
\affiliation{New York University Shanghai, 1555 Century Ave, Pudong, Shanghai 200122, China}

\author{Ebubechukwu O. Ilo-Okeke} 
\affiliation{New York University Shanghai, 1555 Century Ave, Pudong, Shanghai 200122, China}
\affiliation{Department of Physics, School of Science, Federal University of Technology, P. M. B. 1526, Owerri, Imo state 460001, Nigeria}

\author{Valentin Ivannikov}
\affiliation{New York University Shanghai, 1555 Century Ave, Pudong, Shanghai 200122, China} 

\author{Tim Byrnes}
 \email{tim.byrnes@nyu.edu}
	\affiliation{New York University Shanghai, 1555 Century Ave, Pudong, Shanghai 200122, China}
 	\affiliation{State Key Laboratory of Precision Spectroscopy, School of Physical and Material Sciences, East China Normal University, Shanghai 200062, China}
	\affiliation{NYU-ECNU Institute of Physics at NYU Shanghai, 3663 Zhongshan Road North, Shanghai 200062, China}
 \affiliation{Center for Quantum and Topological Systems (CQTS), NYUAD Research Institute, New York University Abu Dhabi, UAE.}
	\affiliation{Department of Physics, New York University, New York, NY 10003, USA}

\date{\today}% It is always \today, today,
    % but any date may be explicitly specified

\begin{abstract}
We study the effects of decoherence in the form of optical phase diffusion, photon loss and gain, and atomic dephasing in entangled atomic ensembles produced via quantum nondemolition (QND) measurements. For the optical decoherence channels, we use the technique of integration within ordered operators (IWOP) to obtain the Kraus operators that describe the decoherence.  We analyze the effect of different decoherence channels on a variety of quantities such as the variances of the spin operators, entanglement and correlation criteria, logarithmic negativity, and the Bell-CHSH inequality. We generally find a smooth decay of correlations and entanglement in the presence of decoherence.  We find that various quantities retain showing non-classical properties under all three types of decoherence, in the short interaction time range. Our results show that such QND measurements are one of the most promising methods for entanglement generation between two Bose-Einstein condensates. 
\end{abstract}

\maketitle

\section{Introduction}
% Macroscopic quantum effects
Quantum entanglement \cite{RevModPhys.81.865,2010Quantum,PhysRevA.72.034305,2007Creating} is one of the central concepts in quantum mechanics, and in quantum information science is considered to be 
an indispensable resource to perform tasks that exceed the performance of classical physics 
\cite{MA201189,70d3af4f8053428eb11a0c4990c219d8,2014On,gerry_knight_2004,2014Quantum,PhysRevX.4.041025,buluta2009quantum,georgescu2014quantum,byrnes2007quantum}. Quantum entanglement has been generally associated with the microscopic world, but with the progress of technology, quantum entanglement has been observed increasingly in the macroscopic world \cite{kunkel2018spatially,fadel2018spatial,lange2018entanglement,2013Entanglement}  Spin squeezing is a prime example of a practical application of such macroscopic entanglement \cite{PhysRevD.4.1925,Walls1983Squeezed,PhysRevA.47.5138,2014Generation}, where the quantum noise of an observable is suppressed below the standard quantum limit \cite{sorensen2001many,kunkel2018spatially,fadel2018spatial,kuzmich2000generation,2008Squeezing,you2017multiparameter}.  Optical systems were the first system where such squeezed states were realized in experiment \cite{slusher1985observation,breitenbach1997measurement,wu1986generation}. One well-studied way that such squeezed states can be produced is through nonlinear effects such as the Kerr effect \cite{2011Introduction,PhysRevA.49.2033}. There are however other ways in which squeezing can be generated, including quantum nondemolition (QND) measurements \cite{hammerer2010quantum}, adiabatic transitions \cite{bec5}, state-dependent forces \cite{ treutlein2006microwave}, Rydberg excitations \cite{bec7}, splitting a single squeezed BEC \cite{bec6},
cavity QED \cite{2013Entanglement,2014Entanglement,bec5,hussain2014,abdelrahman2014coherent}.  Squeezed states have also been obtained in some other systems such as atomic ensembles \cite{gross2012spin,byrnes2020quantum}.  Using QND measurements squeezed states have been generated in atomic ensembles \cite{PhysRevLett.100.103601,2009Implementation,sewell2012magnetic,doi:10.1126/science.aaf3397,PhysRevLett.116.093602}. There are some other systems like mechanical resonators where squeezing has also been experimentally realized \cite{WOS:000433175900001,WOS:000543526500011,WOS:000295255300018,WOS:000428113200003,WOS:000377248800002,WOS:000323895600001}.

Squeezed states are a type of multipartite entangled state \cite{radhakrishnan2020multipartite} where the collective degrees of freedom, rather than the individual identity of the underlying particles is the quantity of interest.  In the context of Bell correlations, however, the identity and the location of the particules is a crucial property and underlies quantification of the non-local correlations using the Clauser-Horne-Shimony-Holt (CHSH) inequality\cite{clauser1969proposed}.  For multipartite states, the two-mode squeezed state is one of the most important type of squeezed states which exhibit Einstein-Podolsky-Rosen  (EPR) correlations. Such states can also be produced in other systems in addition to optical systems, where the first demonstrations of atomic ensembles were pioneered by  Polzik and co-workers \cite{krauter2011entanglement,RevModPhys.82.1041,julsgaard2001experimental,kong_measurement-induced_2020,PhysRevLett.92.030407,PhysRevLett.85.5643,PhysRevLett.85.5639}. In these experiments, two spatially separated Cesium gas clouds \cite{krauter2013deterministic} were entangled using QND measurements \cite{1999Takahashi,2009Phase,higbie2005,PhysRevLett.92.030407,PhysRevA.94.013617,ilo2014theory}.  Applications of the entanglement, such as quantum teleportation for continuous variables \cite{braunstein2005quantum} was successfully
realized \cite{doi:10.1126/science.1253512,krauter2011entanglement,Sherson_2008}.  For Bose-Einstein condensates (BECs), the creation of many-particle entanglement in single BECs has been studied in various contexts \cite{Gross_2012,2001Many,Augusto2014Fisher,PhysRevLett.86.4431}.
Entanglement in different spatial regions of a single BEC has been observed \cite{kunkel2018spatially,fadel2018spatial,lange2018entanglement,2013Entanglement}. However, the entanglement between two  spatially distinct BECs has not been realized in experiments to date. Numerous entangling schemes between BECs have been proposed, ranging from methods using adiabatic transitions \cite{bec5}, state-dependent forces 
\cite{treutlein2006microwave}, Rydberg excitations \cite{bec7}, splitting a single squeezed BEC \cite{bec6}, and 
cavity QED \cite{2013Entanglement,2014Entanglement,bec5,hussain2014,abdelrahman2014coherent}. Generating entanglement is elementary to various quantum information tasks
based on atomic ensembles, such as quantum teleportation \cite{pyrkov2014quantum,pyrkov2014full}, remote state preparation,  clock synchronization \cite{ilo2018remote,PhysRevA.103.062417},  and quantum computing \cite{byrnes2012macroscopic,BYRNES2015102,abdelrahman2014coherent}.

\begin{figure}[t]
\includegraphics[width=\linewidth]{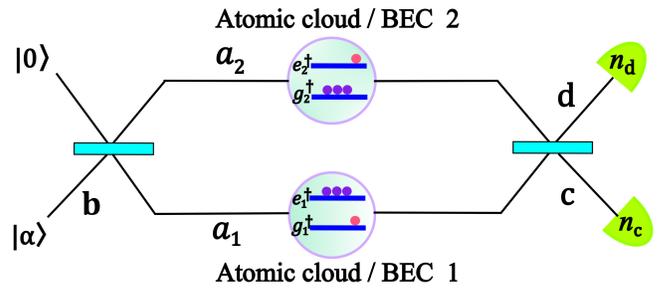}
\caption{\label{Experimental_scheme} Experimental scheme for generating entanglement between two atomic ensembles via QND measurements. A coherent light pulse $|\alpha\rangle $ enters the first beam splitter of a Mach-Zehnder interferometer. Each atomic ensemble or BEC has two relevant internal energy states that are populated by atoms, with
corresponding bosonic annihilation operators $g_j,e_j$, where $j \in \{1,2 \}$ labels each atomic ensemble or BEC. The optical mode is split into two modes labeled by $ a_j $, $j \in \{1,2 \}$ after the first beam splitter. Then each optical mode interacts with the atoms via the QND Hamiltonian. After the second beam splitter, the photons are detected with counts $ n_c $ and $ n_d $. 
 }
\end{figure}

% In this paper
In this paper, we calculate the effects of decoherence in QND measurement-based entanglement generation between atomic ensembles \cite{AristizabalZuluaga2021QuantumNM,Gao_2022,PhysRevA.105.022443}. Previously, a protocol generating entanglement between BECs based on QND measurement induced entanglement was theoretically analyzed \cite{AristizabalZuluaga2021QuantumNM}. The key difference of this theory of QND measurements in comparison to past theories \cite{kuzmich2000generation,julsgaard2001experimental,PhysRevLett.85.5643,serafin2021nuclear,tsang2012evading} is that the QND dynamics is exactly solved such that Holstein-Primakoff (HP) approximation is not used, and non-Gaussian effects can be examined. Another difference is the geometry of the light that is used (see Fig. \ref{Experimental_scheme}) where a Mach-Zehnder configuration is used, instead of the sequential configuration of \cite{pettersson2017light,julsgaard2001experimental}.  This is advantageous in the case where there are many ensembles, as the entanglement can be generated in a scalable all-to-all configuration. 
We investigate the effect of optical phase diffusion, loss and gain of the light, and atomic dephasing  on the entangled state that is generated by the QND scheme.  For the optical decoherence channels, we use the integral within the ordered operator (IWOP) approach \cite{Wang2013StatisticalPA,PhysRevA.82.043842} to analytically evaluate the decoherence effects. This allows for a direct derivation of the Kraus operators for the decoherence channels, giving a powerful way of evaluating the effects.  We note that we have previously performed a study of decoherence on QND-induced entangled states of BECs \cite{Gao_2022}.  This work differs from our previous work in that we analyze phase diffusion and incoherent gain which was not done before.  For the atomic dephasing, we also analyze several different quantities that were not calculated before, such as the violation of CHSH inequality \cite{PhysRevLett.23.880,PhysRevA.98.052115}.

%table of contents
The structure of this paper is as follows. Sec. \ref{ii} briefly reviews the QND entangling protocol, defining the physical model and describing the basic system we are working on. Sec. \ref{phase diffusion} examines the effect of optical phase diffusion on the optical modes involved in the QND measurement.  In Sec. \ref{low-order laser channel}, the effects of photonic loss and gain are examined.  In Sec. \ref{dephasing add}, we analyze the effects of atomic dephasing and show its effect on the Wineland squeezing parameter and  the Bell-CHSH inequality.   Finally,  the results are then summarized in Sec. \ref{conclusion}.

\section{QND Entangling scheme} \label{ii}
In this section, we briefly summarize the theory for QND-induced  entanglement generation between two atomic ensembles or BECs \cite{AristizabalZuluaga2021QuantumNM}.

\subsection{QND Measurement Model and Hamiltonian}\label{ii.A}

We first give a brief description of the QND entangling scheme as shown in Fig. \ref{Experimental_scheme}. Two atomic ensembles or BECs are placed in well-separated traps such as separate magnetic traps on an atom chip or two optical dipole traps \cite{reichel2011atom,whitlock2009t,abdelrahman2014coherent}. The atoms have two internal energy states that are populated. For instance, a suitable choice of the two internal states for  $^{87} \text{Rb}$ could be two hyperfine ground states ($F = 1,m_F =-1$ and $ F = 2,m_F=+1$) of an atom.  These have bosonic annihilation operators $g_j,e_j$ respectively, where $j \in \{1,2 \}$ is used to distinguish the two atomic ensembles or BECs. 

We introduce the collective spin using the Schwinger boson operators, which are defined by the form as follows 
\begin{align}
\label{schwinger}
S^x_j & = e^{\dag}_j g_j + g^{\dag}_j e_j, \nonumber \\
S^y_j & = -i e^{\dag}_j g_j + i g^{\dag}_j e_j, \nonumber \\
S^z_j & = e^{\dag}_j e_j - g^{\dag}_j g_j . 
\end{align}
The commutation relations for Schwinger boson operators are $ [S^l, S^m] = 2i \epsilon_{lmn} S^n $, where $ \epsilon_{lmn} $ is the completely anti-symmetric Levi-Civita tensor with $ l,m,n \in \{ x,y,z \} $. The spin operators (\ref{schwinger}) most directly describe atoms within a BEC, where the atoms are indistinguishable bosons.  The algebra of spin operators for an ensemble of distinguishable atoms is equivalent as long as the state of the system remains in the subspace that is symmetric under particle interchange \cite{byrnes2020quantum}. Hence we will use the above formalism to equally describe an atomic ensemble as well as a BEC.

We define the spin coherent state of the atomic ensemble or BECs to be
\begin{align}
|\theta, \phi\rangle\rangle_j &\equiv \frac{1}{\sqrt{N!}} \left( e^\dag_j \cos \frac{\theta}{2} 
 + e^{i\phi} g^\dag_j \sin \frac{\theta}{2} \right)^N | \text{vac} \rangle \nonumber \\
& = \sum_{k=0}^{N}\sqrt{\binom{N}{k}}\cos^{k}(\frac{\theta}{2})\sin^{N-k}(\frac{\theta}{2})e^{i(N-k)\phi } |k\rangle_j .
\label{coherent state expression}
\end{align}
The direction of the polarization is given by the   angles  $ 0 \leq \theta \leq \pi$, $ -\pi \leq \phi \leq \pi$ on the Bloch sphere. The initial state of atoms is taken to be spins polarized in the $ x $-direction
\begin{align}
    |\frac{\pi}{2},0 \rangle \rangle_1  |\frac{\pi}{2},0 \rangle \rangle_2 .  
       \label{initialatom}
\end{align}
Meanwhile, the Fock states on the $j$th atomic ensemble are defined as
\begin{align}
{|k\rangle}_{j}=\frac{\left(e_{j}^{\dagger}\right)^{k}\left(g_{j}^{\dagger}\right)^{N-k}}{\sqrt{k !(N-k) !}}|\operatorname{vac}\rangle . 
\label{fock state}
\end{align}
The initial state of optical mode $b$ is a coherent light that could be defined as 
\begin{equation}
\label{light coherent state}
|\alpha \rangle_b \equiv e^{ -|\alpha|^2 /2 } e^{\alpha b^\dagger} | \text{vac} \rangle. 
\end{equation}
The coherent light enters the first beam splitter and is split into two modes labeled by $a_j$, where $j\in \left \{ 1,2 \right \} $, evolving to the state
\begin{align}
   |\frac{\alpha}{\sqrt{2}} \rangle_1  |\frac{\alpha}{\sqrt{2}} \rangle_2 .  
   \label{initialphoton}
\end{align}
Meanwhile, the initial state of the atomic ensembles is prepared in the $S^x$ direction
\begin{align}
|\Psi_0 \rangle &= \left. |\frac{\pi}{2},0\rangle \rangle_1 \right.\left. |\frac{\pi}{2},0\rangle \rangle_2 \right. \nonumber \\ & = |k_x=N \rangle_1|k_x=N \rangle_2  . 
\label{becinitial}
\end{align}

Each optical mode interacts with a BEC that is separated in two arms of a Mach-Zehnder interferometer via the QND Hamiltonian  
\begin{align}
\label{Hamilton new}
H=\frac{\hbar \Omega}{2}\left(S_{1}^{z}-S_{2}^{z}\right) 
J^{z} .
\end{align}
This entangles the light and the atomic spin degrees of freedom. 

%Using traditional HP approximation, the initial coherent light must be polarized in a specific direction, while this is not necessary in our case with an exact solving method taken.

\subsection{Entangled wavefunction after QND measurement  }\label{entanglement wavefunction}

Evolving the initial states (\ref{initialatom}) and (\ref{initialphoton}) with the Hamiltonian (\ref{Hamilton new}), and evolving the photons through the second beamsplitter gives the state \cite{AristizabalZuluaga2021QuantumNM}
\begin{align}
    |\psi(\tau) \rangle = & 
    \frac{1}{2^{N}} \sum_{k_{1}, k_{2}=0}^{N}
\sqrt{\binom{N}{k_{1}} \binom{N}{k_{2}}}  |k_1, k_2 \rangle 
 \nonumber \\
 & \times\left|\alpha \cos \left(k_{1}-k_{2}\right)
\tau\right\rangle_{c} \left|-i \alpha \sin \left(k_{1}-k_{2}\right) \tau\right\rangle_{d}  .
    \label{preprojwavefn}
\end{align}
Then projecting the light on Fock states gives rise to a modified atomic wavefunction \cite{AristizabalZuluaga2021QuantumNM}
\begin{align}
|{\psi}_{n_c n_d}(\tau)\rangle &= \frac{1}{\sqrt{\cal N}} \sum_{k_1, k_2 =0}^N \sqrt{{{N}\choose{k_1}}{{N}\choose{k_2}}} \nonumber\\
&\times C_{n_c n_d}^\alpha [(k_1 - k_2) \tau] |k_1,k_2 \rangle .
\label{eq:FinaleUnnormalized}
\end{align}
Here the dimensionless time is defined by $ \tau =\Omega t$.  
The $C$-function is defined by
\begin{align}
C_{n_c n_d}^\alpha (\chi) \equiv \frac{\alpha^{n_c+n_d} e^{-|\alpha|^2/2}}{ \sqrt{n_c ! n_d !} } \cos^{n_c} \chi \sin ^{n_d} \chi , 
\label{cfuncdef}
\end{align}
and the normalization factor ${\cal N}$ is 
\begin{align}
{\cal N} = \sum_{k_1, k_2 =0}^N {{N}\choose{k_1}}{{N}\choose{k_2}} | C_{n_c n_d}[(k_1 - k_2) \tau] |^2 .  
\end{align}

The most likely photon counts that are measured are distributed around $n_{c} +n_{d}\approx |\alpha|^{2} $, due to photon number conservation.  The $ C$-function takes the form of a Gaussian and can be approximated in the short interaction time regime as \cite{kondappan2022imaginary}   
\begin{align}
    C_{n_c n_d}( \chi) \propto \exp
    \left( - \frac{ [| \chi | - \frac{1}{2} \arccos ( \frac{n_c - n_d}{n_c + n_d} ) ]^2}{2 \sigma_{n_c n_d}^2 } \right) , 
\end{align}
where 
\begin{align}
\sigma_{n_c n_d} \approx \frac{1}{\sqrt{n_c + n_d}} . 
\end{align}
For outcomes $ n_c \gg n_d $ and large photon numbers $ |\alpha|^2 \gg 1 $, the state is then well approximated by
\begin{align}
|{\psi}_{n_c \gg n_d}(\tau)\rangle \approx \left(\frac{4}{\pi N}\right)^{1 / 4} \sum_{k=0}^{N} e^{-\frac{2}{N}\left(k-\frac{N}{2}\right)^{2}}|k\rangle|k\rangle . 
\end{align}
This is an entangled state due to the correlations in Fock states that are present.

\section{Optical phase diffusion}
\label{phase diffusion}

The first type of decoherence that we will analyze is phase diffusion for the optical modes. During the QND scheme as shown in Fig. \ref{Experimental_scheme}, the optical modes may undergo phase diffusion.  The state prior to photon detection thus becomes a mixed state, which results in a source of decoherence  for the final atomic wavefunction.  We will use the IWOP technique to obtain the exact density matrix after applying the phase diffusion channel. We illustrate the effects of the decoherence on various quantities such as logarithmic negativity, variances of spin correlators,  probability distributions, correlation-based criteria, and Bell-CHSH inequalities.

\subsection{Phase diffusion master equation}
\label{phase diffusion master dection}

An optical mode $ c $ undergoing phase diffusion is described by the master equation \cite{liu2014master,Carmichael}
\begin{align}
 \frac{d}{d t} \rho=- \kappa \left(c^{\dagger} c \rho+\rho c c^{\dagger}-c \rho c^{\dagger}-c^{\dagger} \rho c\right) ,
 \label{phase diffusion master equation}
\end{align}
where $\kappa$ is the phase diffusion rate. We use the methods of Ref. \cite{PhysRevA.54.958} to exactly  solve the master equations.  This is done by transforming the density operators into ordinary functional equations by using the representation of thermal entanglement states. Exact solutions to the master equations are obtained using the technique of integration within ordered operators
(IWOP). The density matrix including the effects of phase diffusion is given as \cite{doi:10.1142/S0217984908017072}
\begin{align}
\rho(t) = \sum_{m, n=0}^{\infty} M_{m,n}^{\text{PD}} (t) \rho (0) {M_{m, n}^{\text{PD}}}^{\dagger} (t)
\end{align}
where the Kraus operators for phase diffusion are defined as 
\begin{align}
M_{m, n}^{\text{PD}} (t) =\sqrt{\frac{1}{m ! n !} \frac{(\kappa t)^{m+n}}{(\kappa t+1)^{m+n+1}}} c^{\dagger m}\left(\frac{1}{1+\kappa t}\right)^{c^{\dagger} c} c^{n}  .
\label{phase kraus}
\end{align}
Here, $\rho(0)$ is the initial density matrix without decoherence. 

In our case, phase diffusion is applied at the end of the QND entanglement preparation process. We note that due to the two modes $ c, d$ there are two sets of Kraus operators that need to be applied.  The initial density matrix in our case is therefore
\begin{align}
\rho_{0}&= | \psi( \tau) \rangle \langle \psi( \tau) | ,
 \label{initial density matrix}
\end{align}
where $ | \psi( \tau) \rangle  $ is the state as defined in  (\ref{preprojwavefn}).  Applying the phase diffusion and performing the photonic measurements gives the density matrix
\begin{align}
 \rho_{\text{PD}} = & \langle n_c | \langle n_d | \sum_{m,n,m',n'=0}^{\infty} M_{m, n}^{\text{PD},c}  (t)  M_{m', n'}^{\text{PD},d}  (t) \nonumber \\
& \times \rho_{0} {M_{m', n'}^{\text{PD},d}}^{\dagger} (t) 
 {M_{m, n}^{\text{PD},c}}^{\dagger} (t) | n_c \rangle | n_d \rangle  ,
\label{phase final density}
\end{align}
where we have labeled a mode on the Kraus operator with a superscript. 
The explicit evaluated expression is given in Appendix \ref{app:phasediff}. The above expression assumes that the time evolution of the QND interaction $ \tau $ is equal to the time that phase diffusion occurs for in (\ref{phase diffusion master equation}). The rate at which the QND interaction occurs $ \Omega $ and the dephasing rate $ \kappa $ can be adjusted according to the ratio $\tilde{\kappa} = \kappa/\Omega $. Using this density matrix we calculate various quantities of interest as below.

% Therefore, a very important problem is the solution of the master equation \cite{Mufti1993FinitedimensionalMR,Arnoldus:96,PhysRevA.47.5093,PhysRevA.62.013819}. In the past, there have been two main approaches to the master equation. One is to correspond the density operator to various classical functions, such as particle number representation (Q function) coherent state representation (P -representation), or Wigner representation. Another way is to use super operator $\$$ \cite{ArvaloAguilar1998SolutionTT,PhysRevA.67.024101}.

\subsection{Effect of decoherence on various quantities}

\subsubsection{Entanglement}
\label{phase entanglement}

We first examine the effect of phase diffusion on the entanglement between the atomic ensembles.  We use logarithmic negativity to quantify the entanglement, which is an entanglement monotone for mixed states \cite{PhysRevLett.95.090503,PhysRevA.65.032314}
\begin{align}
E=\log _{2}\left\|\rho^{T_{2}}\right\| ,
\label{logarithmic negativity}
\end{align}
where $\rho^{T_2}$ is partial transpose on the second BEC.  We calculate the normalized logarithmic negativity, in relation to the maximum value that is possible between two $ N + 1$ dimensional systems
\begin{align}
E_{\max }=\log_{2}(N+1) . 
\end{align}

\begin{figure}[t]
\centering
\includegraphics[width=\columnwidth]{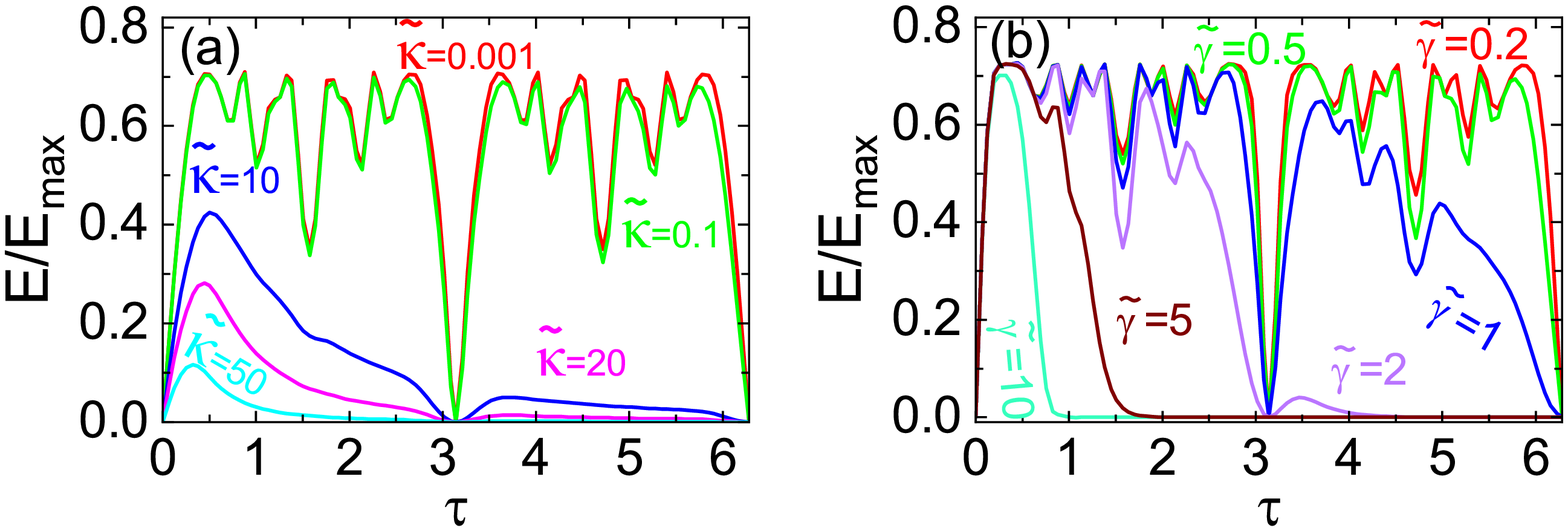}
\caption{Entanglement as quantified by the normalized logarithmic negativity (\ref{logarithmic negativity}) for the state (\ref{preprojwavefn}) in the presence of different kinds of decoherence channels. (a) Entanglement in the phase diffusion channel versus time with decoherence $ \tilde{\kappa} = \kappa/\Omega $ as marked.  (b) Entanglement for the photonic loss/gain channel versus time with decoherence rates as marked. We set the photonic gain rate to be $g/\Omega =0.1$ and show various photon loss rates $ \tilde{\gamma} = \gamma/\Omega $.  
The parameters used in (a) are $N=10,n_c=20,n_d=0,\alpha=\sqrt{20}$ while $N=10,n_c=100,n_d=0,\alpha=10$ for other plots.}
\label{entanglment plots}
\end{figure}

In Fig. \ref{entanglment plots}(a), we show the dependence of the entanglement for various values of the dimensionless phase diffusion rate  $\tilde{\kappa}  $.  We observe that for small values of $ \tilde{\kappa} \le 0.1 $, the entanglement is relatively unaffected for all interaction times. The curve shows a characteristic ``devil's crevasse'' form as observed in previous works \cite{byrnes2013fractality,pettersson2017light}. With increasing $\tilde{\kappa}$, the entanglement shows an exponential decay, and the sharp dips are smoothed out.  The entanglement rises sharply corresponding to the generation of entanglement using the QND scheme within the dephasing time.  For longer times, the entanglement is increasingly affected by the dephasing, due to the longer times that the decoherence has to degrade the state.  Beyond $\tilde{\kappa}\ge 10$, the devil’s crevasse structure is no longer visible.  We note that there are no decoherence effects at times $\tau= n \pi$, due to such times having no entanglement even in the decoherence-free case.  This does not necessarily mean that the state is unaffected since the entanglement is only one aspect of the state and may still affect the state.

%Our one-pulse exact state (\ref{eq:FinaleUnnormalized}) exhibits entanglement between the two BECs. However, in Ref. \cite{AristizabalZuluaga2021QuantumNM} this was only examined under ideal conditions as no forms of decoherence were studied. 

\subsubsection{Probability distribution}
\label{correlation}

We next consider the probability distribution of measuring  state (\ref{phase final density}) in various bases 
\begin{align}
p_{l}(k_1,k_2)=\langle k_1, k_2 |^{(l)} \rho |k_1, k_2\rangle^{(l)}. \label{correlation}
\end{align}
where we use the notation
\begin{align}
| k_1, k_2 \rangle^{(l)} = | k_1 \rangle^{(l)} \otimes  | k_2 \rangle^{(l)}
\end{align}
for $ l \in \{ x, y, z \} $.  The Fock states in the $ x $ and $ y $ bases are defined as 
\begin{align}
|k\rangle^{(x)} &=e^{-i S^{y} \pi / 4}|k\rangle^{(z)} \nonumber \\
|k\rangle^{(y)} &=e^{-i S^{z} \pi / 4} e^{-i S^{y} \pi / 4}|k\rangle^{(z)} .
\label{fock state in various bases}
\end{align}
where the Fock state $ |k\rangle^{(z)} $ is the same as that defined in (\ref{fock state}).  The Fock states in the $ x$ and $ y $-basis can be written in terms of the $ z $-basis Fock states using the relations given in Appendix \ref{app:fockxyz}.

\begin{figure}[t]
\centering
\includegraphics[width=\columnwidth]{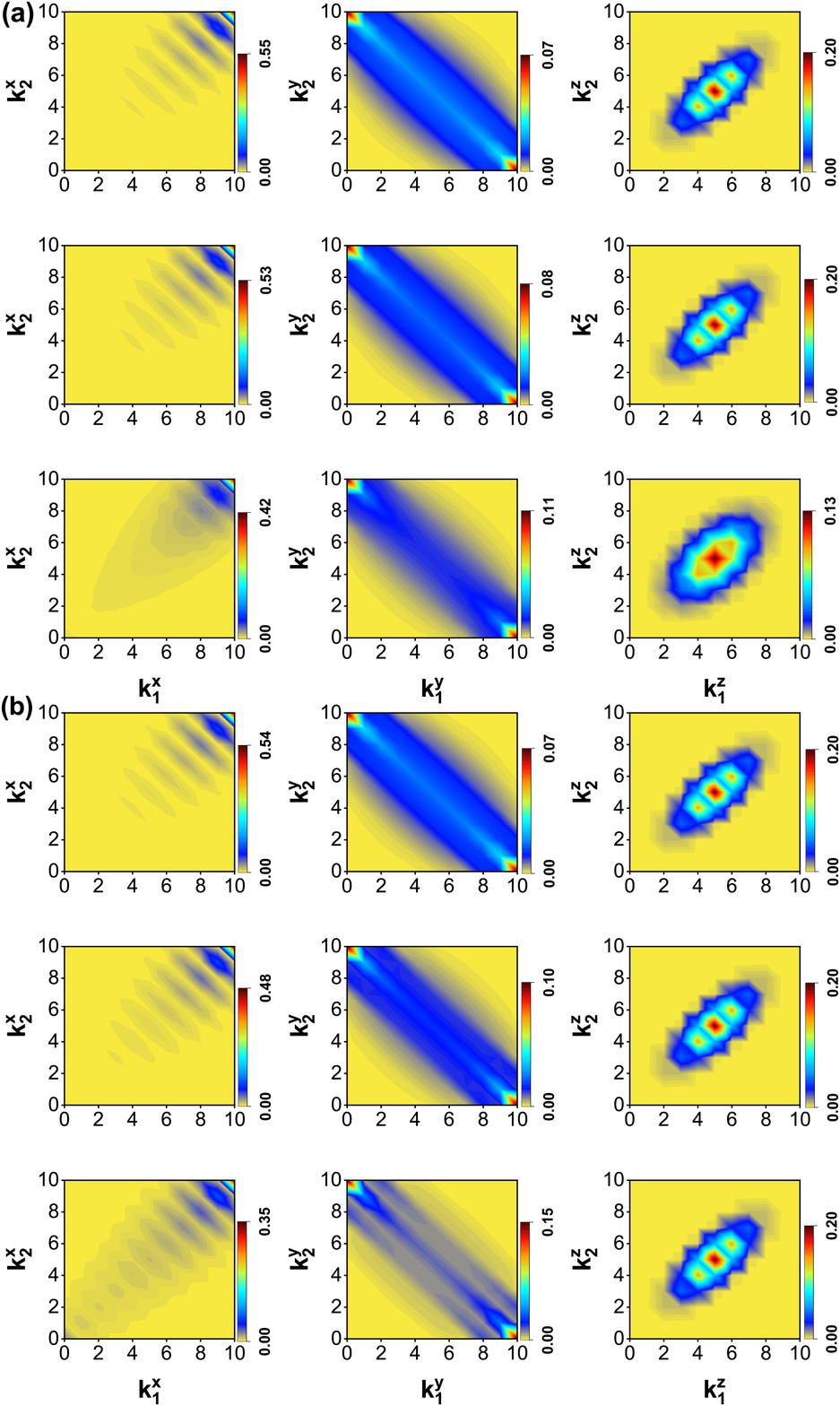}
\caption{Probability distributions after measuring state (\ref{eq:FinaleUnnormalized}) with decoherence in various bases, as defined in (\ref{correlation}).  (a) The phase diffusion channel with dissipative coefficient $\tilde{\kappa} \in \{0,1,20\}$ from top to bottom, for the state (\ref{phase final density}).  (b) The photonic loss/gain channel with loss coefficient $\tilde{\gamma} \in \{0.2,1,10\}$ from top to bottom, and $ g/\Omega=0.1$. The parameters for all plots are $N=10,n_c=100,n_d=0,\tau=0.1$.  }
\label{density1}
\end{figure}

Figure \ref{density1}(a) shows the probabilities with phase diffusion channel for various amounts of decoherence $ \tilde{\kappa} $ at times $ \tau = 0.1$, corresponding to times where two-spin squeezing is visible.  For the decoherence-free case, we see a pattern of correlations for $ l=x,z$ and anti-correlations for $ l = y $.  As the dephasing is increased, the pattern of correlations and anti-correlations are modified. The spin correlations in the $z$-direction are decreased, and the diagonal probability distribution is modified to circular distribution which exhibits no correlations. However, the correlations do not become completely removed even for very large amounts of phase diffusion, such as $ \tilde{\kappa} = 20$.  Interestingly, the $ y$-correlations appear to be improved under phase diffusion, where the width of the anti-correlations is reduced.  We confirm this effect in the variances in the following section.  For the $ x$-correlations, from the probability distribution, there is little visible change under the phase diffusion, except for the removal of interference fringes.  Overall, we observe that even under very strong phase diffusion, the correlations remain relatively unaffected for the probability distributions that are shown in Fig. \ref{density1}(a).  This is likely due to the relatively short interaction time that is examined, where the type of entangled state is relatively robust in the presence of decoherence \cite{byrnes2013fractality}.

\subsubsection{Variances}
\label{phasevariance}

\begin{figure}[t]
\centering
\includegraphics[width=\columnwidth]{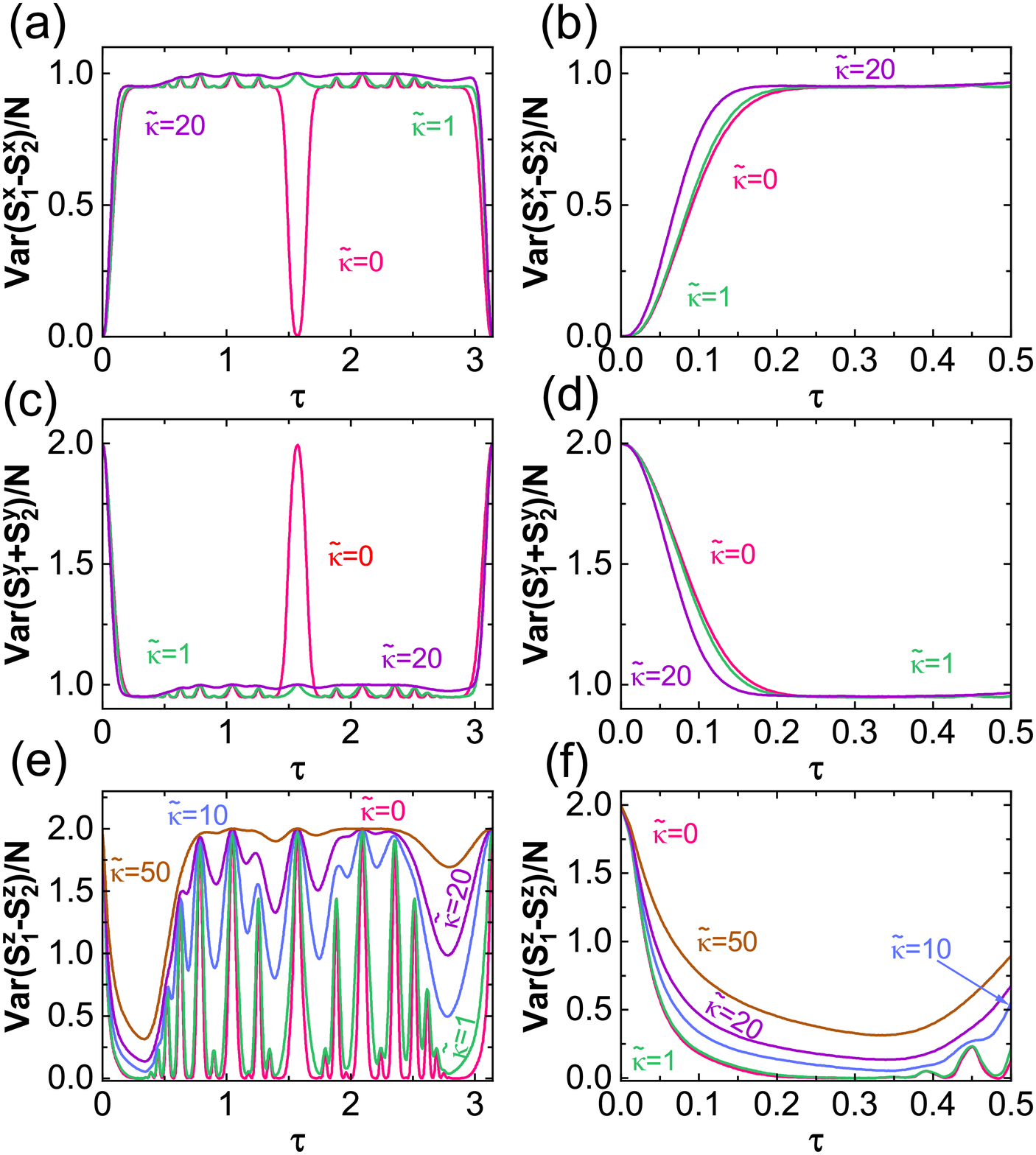}
\caption{The variance of the state (\ref{phase final density}) with  phase diffusion decoherence for the operators $S_1^x-S_2^x,S_1^y+S_2^y$ and $S_1^z-S_2^z$. Variances are plotted for (a)(c)(e) long time scales; (d)(e)(f) zoomed into short time scales. The parameters are $N=10,n_c=100,n_d=0,\alpha=\sqrt{n_c}=10$. }
\label{kuosanfangcha}
\end{figure}

Variances of spin correlators $ S^x_1 - S^x_2 $, $ S^y_1 + S^y_2 $, and $ S^z_1 - S^z_2 $ under the phase diffusion channel are shown as a function of interaction time in Fig. \ref{kuosanfangcha}. We choose these observables since they are expected to show squeezing effects.  Comparing the $\tilde{\kappa}=1$ with the decoherence-free case of $ \tilde{\kappa} = 0 $, we observe that there is very little difference for all spin correlators at most evolution times. The largest differences arise at times  $ \tau =\pi $, where the state involves entangled Schrodinger cat states \cite{Gao_2022}, which are particularly susceptible to decoherence. The variance of $S_1^x-S_2^x$ remains large, whereas this drops to zero in the decoherence-free case.  Similarly, the variance of $S_1^y+S_2^y$ in the decoherence-free case reaches a large value at $ \tau =\pi $ but is reduced when phase diffusion is introduced.  Meanwhile, the variance of $S_1^z-S_2^z$ is relatively affected at all times when comparing the $ \tilde{\kappa} = 1$ and decoherence-free cases.  However, as the phase diffusion rate is further increased, generally the variance of  $S_1^z-S_2^z$ is increased, which can be understood as the loss of correlations.  Meanwhile, the $S_1^x-S_2^x$ and $S_1^y+S_2^y$ correlations are relatively unaffected with a further increase of phase diffusion beyond $ \tilde{\kappa} = 1$.

\subsubsection{Correlation-based criteria}
\label{entanglement criterion}

While logarithmic negativity is a well-established method for detecting entanglement, in an experimental setting it requires full tomography of the quantum state which has a large overhead and may not be practically feasible in many cases.  A preferable approach is to detect entanglement with low-order spin expectation values, which are more readily measured and are relatively insensitive to imperfect spin measurements.  Here we describe and calculate three correlation-based methods to detect entanglement and EPR steering.  

The first correlation-based entanglement criterion is the Wineland squeezing parameter
\begin{align}
\xi & =\frac{{2(\Delta S_{\bot }^{2})}_{\text{min} }}
{\left | \left \langle S  \right \rangle  \right | }  \nonumber \\
& \geq 1 \hspace{1cm} \text{(separable)}
\label{wineland squeezing}
\end{align}
where $S_{\bot }$ is any component perpendicular to the average total spin and $\Delta S_{\bot }^{2}$ is minimum fluctuation in the direction of the vertical component. The coefficient $\xi$ corresponds to the degree of squeezing. A smaller value means a better degree of squeezing. The inequality (\ref{wineland squeezing}) is derived assuming a separable state, hence $ \xi < 1 $ signals the presence of entanglement.

In a previous study \cite{AristizabalZuluaga2021QuantumNM}, it was found that the Hofmann-Takeuchi criterion \cite{PhysRevA.68.032103} is one of the most effective spin correlation-based detection methods for entanglement. In our case, it is defined as
\begin{align}
{\cal {C}}_{\text {ent }} & \equiv \frac{\operatorname{Var}\left(S_{1}^{x}-S_{2}^{x}\right)+\operatorname{Var}\left(S_{1}^{y}+S_{2}^{y}\right)+\operatorname{Var}\left(S_{1}^{z}-S_{2}^{z}\right)}{4 N} \nonumber \\
& \geq 1.  \hspace{1cm} \text{(separable)}
\label{HT criteria}
\end{align}
Again, the inequality (\ref{HT criteria}) is derived assuming separable states, hence detecting $ {\cal {C}}_{\text {ent }} < 1 $ signals the presence of entanglement.

Finally, EPR steering is a subclass of entangled states where one party can nonlocally affect the other party's state through measurements \cite{wiseman2007steering,2018Demonstration,ma2019operational}.  A correlated-based inequality that detects EPR steering is \cite{RevModPhys.81.1727}
\begin{align}
{\cal{C}}_{\text {steer }}^{1 \rightarrow 2} & \equiv \frac{\operatorname{Var}\left(S_{1}^{y}+S_{2}^{y}\right) \operatorname{Var}\left(S_{1}^{z}-S_{2}^{z}\right)}{\left\langle S_{1}^{x}\right\rangle^{2}} \nonumber \\
& \ge 1 \hspace{1cm} \text{(un-steerable)}
\label{EPR steering}
\end{align}
A violation of the inequality (\ref{EPR steering}) implies the existence of EPR steering from BEC 1 to BEC 2.

\begin{figure}[t]
\centering
\includegraphics[width=\columnwidth]{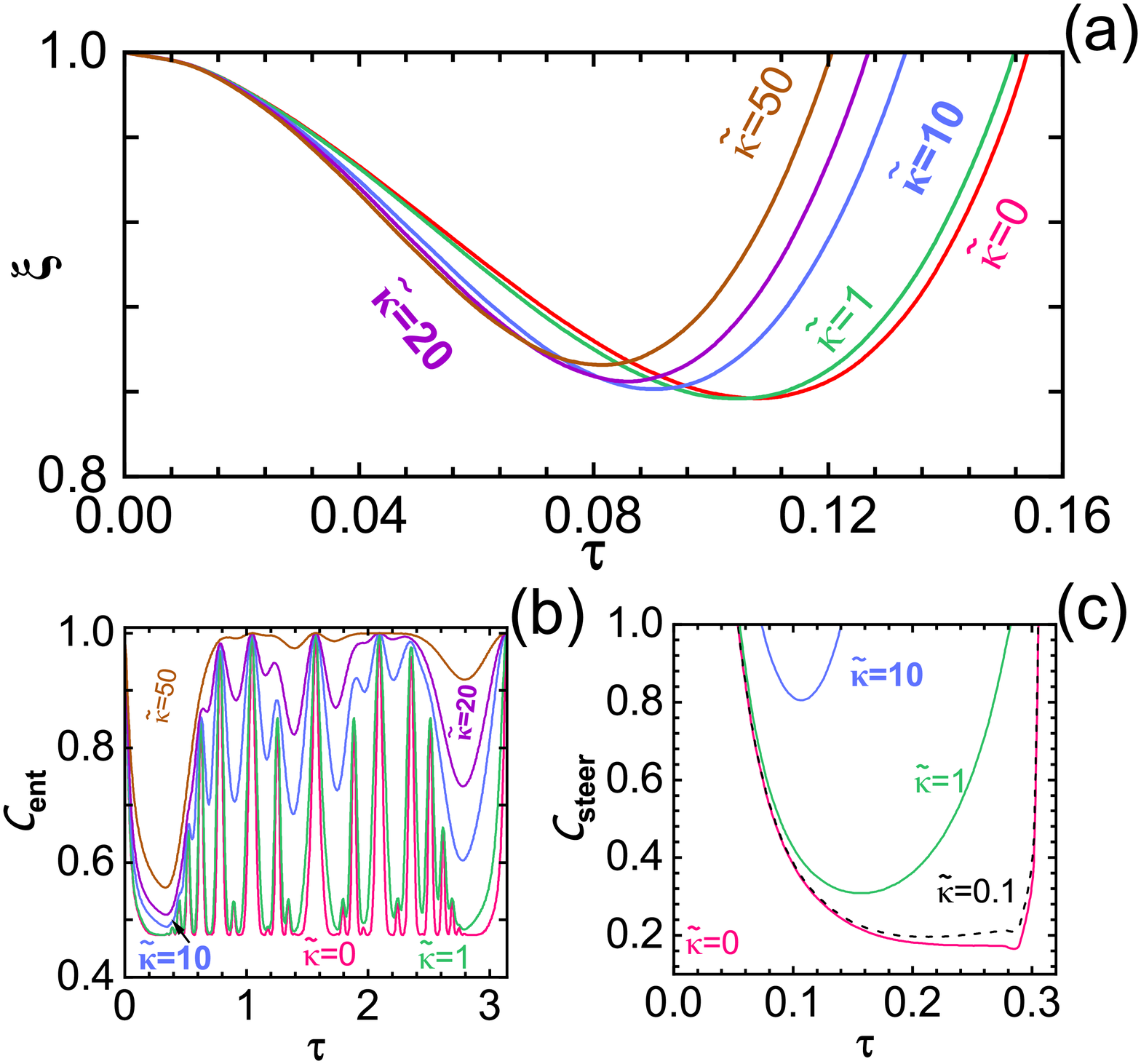}
\caption{Detection of entanglement and EPR-steering using correlation-based criteria. (a) The Wineland squeezing criterion (\ref{wineland squeezing}) for the state (\ref{phase final density}) with phase diffusion decoherence. (b) The Hofmann-Takeuchi criterion (\ref{HT criteria}) for the state (\ref{phase final density}) with phase diffusion decoherence.  (c) The EPR steering criterion (\ref{EPR steering}) for the state (\ref{phase final density}) with phase diffusion decoherence. The parameters used here are $N=10,n_c=100,n_d=0,\alpha=\sqrt{n_c+n_d}=10$. The dissipative coefficient is as marked. We set $ \Omega = 1 $ such that the decoherence rate is in units of $ \Omega $, and the time units are $ 1/\Omega $.}
\label{kuosanentanglment criteria}
\end{figure}

In Fig.\ref{kuosanentanglment criteria}, we compare the above three different correlation-based criteria. Fig.\ref{kuosanentanglment criteria}(a) shows Wineland squeezing parameter in the region where entanglement is shown. The time region of sustained squeezing for which $\xi < 1$ is reduced  as the phase diffusion rate $\tilde{\kappa}$ is increased. This is consistent with the logarithmic negativity calculations of Fig. \ref{entanglment plots}(a), where the phase diffusion has the effect of reducing entanglement. Now turning to the Hofmann-Takeuchi criterion, we find it can detect entanglement for a wide range of times except for some particular points where ${\cal {C}}_{\text {ent }}\approx 1 $.  As with the Wineland squeezing criterion, generally, the degree of violation of the inequality (\ref{HT criteria}) is reduced with an increasing phase diffusion rate.  It however remains a powerful method to detect entanglement, and in comparison to the Wineland squeezing criterion, it detects entanglement in a much wider range.  

Finally, for the EPR steering criterion, the effect of phase diffusion is to decrease the region where violation of the inequality (\ref{EPR steering}) occurs.  As would be expected, for EPR steering it is generally more difficult to find a violating region, which we attribute to the fact that EPR steerable states are a more specialized class of entangled states.  Beyond $ \tilde{\kappa} > 10 $, the criterion fails to detect any entangled states, while the Wineland squeezing and the Hofmann-Takeuchi criterion continue to detect entangled states.

\subsubsection{Bell-CHSH inequality}
\label{Bell-chsh for phase}

\begin{figure}[t]
\centering
\includegraphics[width=\columnwidth]{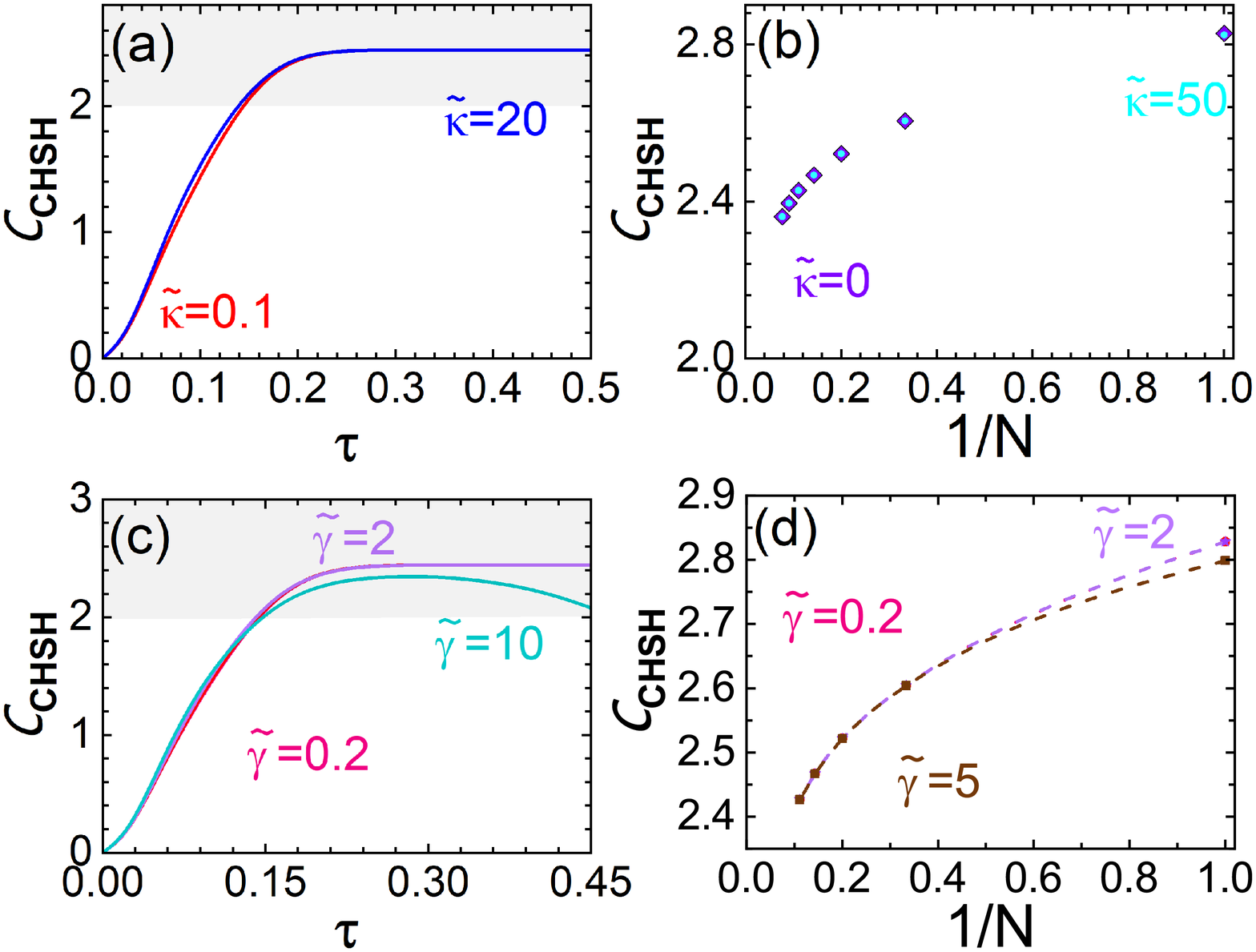}
\caption{Bell-CHSH correlations for the state (\ref{eq:FinaleUnnormalized}) with two types of optical deocherence. (a) Time dependence of (\ref{bell-chsh}) for phase diffused state (\ref{phase final density}) for $N=5 $ and $ \theta_{B}=0.37 $.  (b) Optimal values of (\ref{bell-chsh}) concerning $\tau$ and $\theta_{B}$ for various $N$ with phase diffused state (\ref{phase final density}). 
(c) Time dependence of (\ref{bell-chsh}) for the state (\ref{rholg}) which has undergone photon loss and gain. (d) Optimal values of (\ref{bell-chsh}) concerning $\tau$ and $\theta_{B}$ for various $N$ under photon loss and gain (\ref{rholg}). \  Shaded regions indicate the regions where there is a violation of the CHSH inequality.  The common parameters are $ n_c=100, n_d=0, \alpha=10$.}
\label{allchsh}
\end{figure}

In Ref.\cite{juanQND}, it was shown that the Bell-CHSH inequality can be violated using the state (\ref{eq:FinaleUnnormalized}).   The form of the CHSH inequality that was used, reads as
\begin{align}
\mathcal{C}_{\text{CHSH}} \equiv & \Big| \left\langle M_{1}^{(1)} M_{2}^{(1)}\right\rangle+\left\langle M_{1}^{(1)} M_{2}^{(2)}
\right\rangle \nonumber \\
&-\left\langle M_{1}^{(2)} M_{2}^{(1)}\right\rangle+\left\langle M_{1}^{(2)} M_{2}^{(2)}
\right\rangle \Big| \nonumber \\
\le  &  2 \hspace{1cm} \text{(local HV)},
\label{bell-chsh}
\end{align}
which is valid for a local hidden variable (HV) theory and 
\begin{align}
\nonumber
    M_1^{(1)} &= \text{sgn}( S^{z}_1) \\ \nonumber
    M_1^{(2)} &= \text{sgn}(S^{z}_1 \cos \theta_B  +S^{y}_1 \sin  \theta_B)\\ \nonumber
    M_2^{(1)} &= \text{sgn}(S^{z}_2 \cos \frac{\theta_B}{2}  + S^{y}_2 \sin \frac{\theta_B}{2})  \\ 
    M_2^{(2)} &= \text{sgn}(S^{z}_2 \cos \frac{\theta_B}{2}  - S^{y}_2  \sin \frac{\theta_B}{2}) .
\end{align}
Here, $ \text{sgn}(x) $ takes the sign of the eigenvalue of the operator. Using the sign of spin operators is an experimentally viable observable since it is largely insensitive to atom number fluctuations \cite{Jing_2019}. The superscripts $ ^{(1),(2)}$ are the two measurement choices that can be made on the two atomic ensembles. The angle $ \theta_B $ must be optimized to find the largest violation and was found to be well-approximated by the empirical relation $\theta_B \approx (3.2/N+ 1.7/N^2)/(1+ 2.1/N) $ \cite{juanQND}.  

In Ref. \cite{juanQND}, the Bell-CHSH inequality was studied without considering decoherence effects.  Figure \ref{allchsh}(a) shows the time dependence of the left-hand side of the criterion (\ref{bell-chsh}) with phase channel decoherence for $ N = 5 $. The shaded area indicates the classical region of Bell-CHSH inequality.  Remarkably, we see very little change in the values of $ \mathcal{C}_{\text{CHSH}} $, and above a particular interaction time, CHSH violations are seen despite the presence of strong phase diffusion of the optical modes.  In Fig. \ref{allchsh}(b), the optimal values of (\ref{bell-chsh}) are as a function of $1/N$. We observe the violations occur for all $N$ in the presence of strong phase diffusion. We again find that decoherence has little effect on the optimal Bell-CHSH inequality. While this is a positive from an experimental context, we note that for large ensemble sizes the level of violation is reduced, hence an increased precision of $ \mathcal{C}_{\text{CHSH}} $ will be required to observe any violation.

\section{Photonic Loss and Gain}
\label{low-order laser channel}

We now consider a loss and gain channel for the photons during the QND measurement induced entanglement generation as described in Sec. \ref{entanglement wavefunction}.  Physically, the loss of photons may occur during the transmission of the light through the Mach-Zehnder interferometer and is potentially a major source of decoherence to the scheme.  Photonic gain may also occur where ambient incoherent photons may enter the modes, also acting as a source of decoherence.

\subsection{Photon loss and gain master equation}

The master equation for photonic loss and gain is written as \cite{PhysRevA.4.739,PhysRevA.39.4628,PhysRevA.46.4239}
\begin{align}
\frac{d \rho}{d t}=\gamma \left(2 c \rho c^{\dagger}-c^{\dagger} c \rho-\rho c^{\dagger} c\right)+g\left(2 c^{\dagger} \rho c-c c^{\dagger} \rho-\rho c c^{\dagger}\right) ,
\end{align}
where $ \gamma  $ is the photon loss rate and $ g $ is the photon gain rate. Solving this master equation using the IWOP technique can be written in terms of Kraus operators as \cite{fan2008operator,CHEN2013272} 
\begin{align}
\rho(t) &=\sum_{p, q=0}^{\infty} M_{p, q}^{\text{LG}} (t)  \rho (0)  {M_{p, q}^{\text{LG}}}^{\dagger} (t)
\end{align}
where
\begin{align}
M_{p, q}^{\text{LG}} (t)  &  =  \sqrt{\frac{\gamma^{p} 
g^{q} T_{1}^{p+q} T_{3}  }{p ! q ! T_{2}^{2 q}} }
e^{c^{\dagger} c \ln T_{2}} c^{\dagger q} c^{p} \nonumber \\
T_{1} & =\frac{1-e^{-2(\gamma-g) t}}{\gamma-g e^{-2 (\gamma-g)t }}\nonumber \\ 
T_{2} & =\frac{(\gamma-g) e^{-(\gamma-g) t}}{\gamma-g e^{-2 (\gamma-g)t}}\nonumber \\ 
T_{3} & =\frac{\gamma-g}{\gamma-g e^{-2 (\gamma-g)t }}=1-g T_{1} . 
\end{align}
The above expression reduces to the standard form of the Kraus operator for loss \cite{Gao_2022} when $ g = 0 $, where the probability of the photon loss is $ 1- e^{-2 \gamma t }$.  

Using the above Kraus operator we may derive the effect of photonic loss and gain on the state by applying it to  (\ref{initial density matrix})
\begin{align}
\rho_{\text{LG}} = & \langle n_c | \langle n_d | \sum_{p,q,p',q'=0}^{\infty} M_{p, q}^{\text{LG},c}  (t)  M_{p', q'}^{\text{LG},d}  (t) \nonumber \\
& \times \rho_{0} {M_{p', q'}^{\text{LG},d}}^{\dagger} (t) 
 {M_{p, q}^{\text{LG},c}}^{\dagger} (t) | n_c \rangle | n_d \rangle  .
 \label{rholg}
\end{align}
The evaluated expression is given in Appendix \ref{app:lossgain}.  As with phase diffusion, the above expression assumes that the time evolution of the QND interaction $ \tau $ is equal to the time that phase diffusion occurs for in (\ref{phase diffusion master equation}). We note that this is different to the assumption made in Ref. \cite{Gao_2022}, where the probability of photon loss was set to be a constant, and hence was a time-independent quantity.  The rate at which the QND interaction occurs $ \Omega $ and the loss rate $ \gamma $ can be adjusted according to the ratios $\tilde{\gamma} = \gamma /\Omega $ and $\tilde{g} = g /\Omega $. Using this density matrix we calculate various quantities of interest as below.

\subsection{Effect of decoherence on various quantities}

\subsubsection{Entanglement}
\label{laser entanglement}

In Fig. \ref{entanglment plots}(b), we show the logarithmic negativity for a fixed photon gain rate $ \tilde{g} $ and various rates of loss $ \tilde{\gamma} $. We choose such a regime since it is the most physically relevant regime, where the dominant decoherence mechanism is photon loss.  The constant photon gain accounts for the possibility that there is an incoherent source of photons entering the interferometer, but we assume that this can be suitably controlled such that is a relatively smaller effect.  We again see the characteristic devil’s crevasse structure
to the entanglement as the time is varied.  Generally, the amount of entanglement decreases with longer times due to the assumption here that the time of evolving the photonic loss and gain master equation is equal to the QND interaction time.  This results in a probability of the photon loss that exponentially approaches 1, such that the  entanglement exponentially reduces with time. 

The results shown in Fig. \ref{entanglment plots}(b) are qualitatively different from what was shown in Ref. \cite{Gao_2022}, where there was little difference between the ideal entanglement curves with and without photon loss. We attribute this to the effectively larger photon loss that our current calculations include, compared to Ref. \cite{Gao_2022}.  In Ref. \cite{Gao_2022}, the largest photon loss probability that was considered was 0.95.  In comparison, for example, the photon loss probability for $ \xi = 1 $ at $ \tau = \pi $ is 0.998. Once the photonic population is entirely removed through photon loss, it is clear that the entangling scheme cannot work since the state (\ref{preprojwavefn}) no longer involves any photons and the $ C$-function (\ref{cfuncdef}) no longer produces an amplitude on the states.  We also note that the dependence of the phase diffusion plots in Fig. \ref{entanglment plots}(a), are somewhat different from the photon loss and gain dependence, which tends to maintain high levels of entanglement followed by a sharp drop off.  This is in contrast with  phase diffusion has the effect of decreasing the  overall level of entanglement.  We again interpret this as the QND measurement being relatively robust in the presence of photon loss until all the photons are removed, at which point the protocol no longer is capable of generating entanglement.

\subsubsection{Probability distribution}

Fig. \ref{density1}(b) shows the probability distribution (\ref{correlation}) for the state (\ref{rholg}).  The ideal case without any decoherence is shown in the top row of Fig. \ref{density1}(a). We see that there is very little difference between the ideal case and the case with photon loss $ \tilde{\gamma} =0.2, 1$ and gain $ \tilde{g}=0.1 $.  At very large loss rates $ \tilde{\gamma} =10$, some difference starts to develop in the $ x$ correlations and  $ y $ anti-correlations but the $ z $-correlations remain visually the same.  As with the phase diffusion case, the $ y $ anti-correlations tend to improve slightly for the larger photon loss cases.  We do not have a good interpretation of this effect, but this appears to be a feature of both types of photonic decoherence that the $ y$ anti-correlations are improved.

\subsubsection{Variances}
\label{laser variances }

\begin{figure}[t]
\centering
\includegraphics[width=\columnwidth]{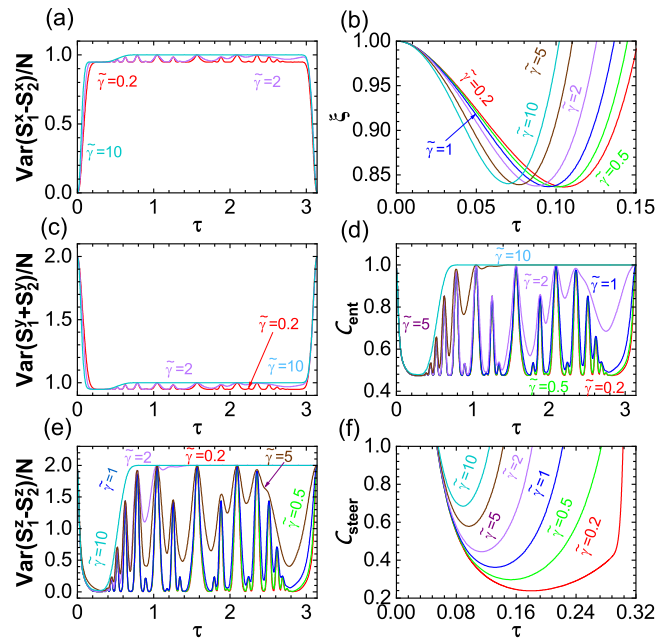}
\caption{Variances and correlation-based criteria for the QND entangled state under photon loss and gain. (a)(b)(c)  Expectation values $S_1^x-S_2^x, S_1^y+S_2^y$ and $S_1^z-S_2^z$; (d) Wineland squeezing; (e) Hofmann-Takeuchi criterion; (f) EPR steering criterion.  The parameters are $N=10,n_c=100,n_d=0,\alpha=\sqrt{n_c}=10$. We set $ \Omega = 1 $ such that $\tilde{\kappa}$ and $g$ is in units of $ \Omega $, and the time units are $ 1/\Omega $.}
\label{laserfangcha}
\end{figure}

Fig. \ref{laserfangcha}(a)(c)(e) shows the variances of the state (\ref{rholg}) for the operators $S_1^x-S_2^x, S_1^y+S_2^y$, and $S_1^z-S_2^z$ respectively for the loss and gain channel. We again consider the case with fixed incoherent photon gain and vary the loss rate.  We see that for the $ x $- and $ y $-basis correlators, there is relatively little change to the dependence despite relatively large amounts of loss present.  Some of the small structure in the variations is removed for $ \tilde{\kappa} = 10 $, but this is a small effect with respect to the overall variance.  For the $ z $-basis correlators, for early times, all values of the variance remain similar, and squeezing is observed.  For longer times, the variance becomes more affected by the loss, and at $ \tilde{\kappa} = 10 $ the variance returns to its initial value. This can be attributed due to the fact that the states at longer evolution times tend to be more sensitive to the decoherence such that it minimizes the correlations.

\subsubsection{Entanglement criteria and Bell-CHSH inequality}
\label{lase critetia and bell}

In Fig.\ref{laserfangcha}(b)(d)(f), we show the three different correlation-based criteria of the Wineland squeezing parameter (\ref{wineland squeezing}), Hofmann-Takeuchi criterion (\ref{HT criteria}), and the EPR steering criterion (\ref{EPR steering}). For the Wineland squeezing parameter, we find a generally similar behavior to phase diffusion, where the region where entanglement is detected tends to reduce with increasing photon loss.  Generally, the criterion only detects entanglement in the short time regime where the initial squeezing of states occurs with respect to the spin variances.  
Fig.\ref{laserfangcha}(d) shows the Hofmann-Takeuchi criterion, which tends to follow the time dependence of the $S_1^z-S_2^z$ variance. This is due to the fact that variances of $S_1^x-S_2^x $ and $S_1^y+S_2^y $ show a perfect symmetry, which means that the sum $\text{Var}(S_1^x-S_2^x)+\text{Var}(S_1^y+S_2^y)$ is a constant. 
We find that this is again a powerful way of detecting entanglement, showing that entanglement is present in the system for nearly all times for most values of $ \tilde{\gamma} $.  The EPR steering criterion in Fig.\ref{laserfangcha}(f) shows again a similar trend where the increased photon loss causes a more limited region where EPR steering is detected. Figure.\ref{allchsh}(c) shows the Bell-CHSH criterion (\ref{bell-chsh}) as a function of time including the effects of photon loss and gain.  Again we see that there is great robustness in observing a violation, even in the presence of strong photon loss.  In Fig.\ref{allchsh}(d) we see the dependence of the level of violation as a function of $ 1/N $.  We again see that for all $ N $ values a Bell violation can be observed and there is very little difference between the levels of violation including photon loss and gain and the ideal case.  The main difficulty in seeing such correlations will then be in observing the Bell's violation for larger atomic ensembles where the level of violation approaches the classical bound of $ {\cal C}_{\text{CHSH}} = 2$.

\section{Atomic Dephasing}\label{dephasing add}

We now examine a third channel of decoherence, atomic dephasing, which directly acts on the atoms to remove coherence in the atomic state.  This may physically arise due to noise sources that the atoms may experience, originating from technical noise in the atomic traps, or due to photonic scattering due to the QND Hamiltonian (\ref{Hamilton new}).  In a previous study \cite{Gao_2022}, the effect of $S^z$ dephasing on the state (\ref{eq:FinaleUnnormalized}) was examined, where quantities such as the expectation values, variances, entanglement criteria, and distribution were  investigated. However, several quantities such as the Wineland criterion on Bell-CHSH inequality were not examined.  We discuss some of the quantities that were not considered in Ref. \cite{Gao_2022} here, to analyze the impact of decoherence.

\subsection{Dephasing master equation}

The master equation for atomic dephasing reads
\begin{align}
\frac{d \rho }{d t }= - \Gamma
\sum_{n=1}^2\left [ \left ( S_{n}^z \right ) ^2\rho -2S_{n}^z\rho S_{n}^z+ \rho \left ( S_{n}^z \right ) ^2 \right ] , 
\label{Markovian}
\end{align}
where $\Gamma$ is the dephasing rate and it has been assumed to be the same for both atomic ensembles. 
 Solving the master equation (\ref{Markovian}) exactly, the density matrix elements evolve according to
\begin{align}
\rho_{\text{D}} = &  \sum_{k_1, k_2, k_1', k_2'}
| k_1, k_2 \rangle \langle k_1', k_2' | \langle k_1, k_2 | \psi_{n_c n_d} (\tau) \rangle  \nonumber \\
& \times \langle \psi_{n_c n_d}  (\tau) | k_1' , k_2' \rangle   e^{-2 \tilde{\Gamma} \tau  [ (k_1 - k_1')^2 + (k_2 - k_2')^2 ]} ,  
\label{rhosol}
\end{align}
where the atomic wavefunction after the photonic measurement is given by (\ref{eq:FinaleUnnormalized}). 
As before, we set the dephasing time to be equal to the QND interaction time.  The relative strengths of the dephasing and the interaction can be adjusted by the dimensionless ratio $ \tilde{\Gamma} = \Gamma/\Omega $.

\subsection{Effect of decoherence on various quantities}

\subsubsection{Wineland squeezing criteria }
\label{wineland}

\begin{figure}[t]
\centering
\includegraphics[width=\columnwidth]{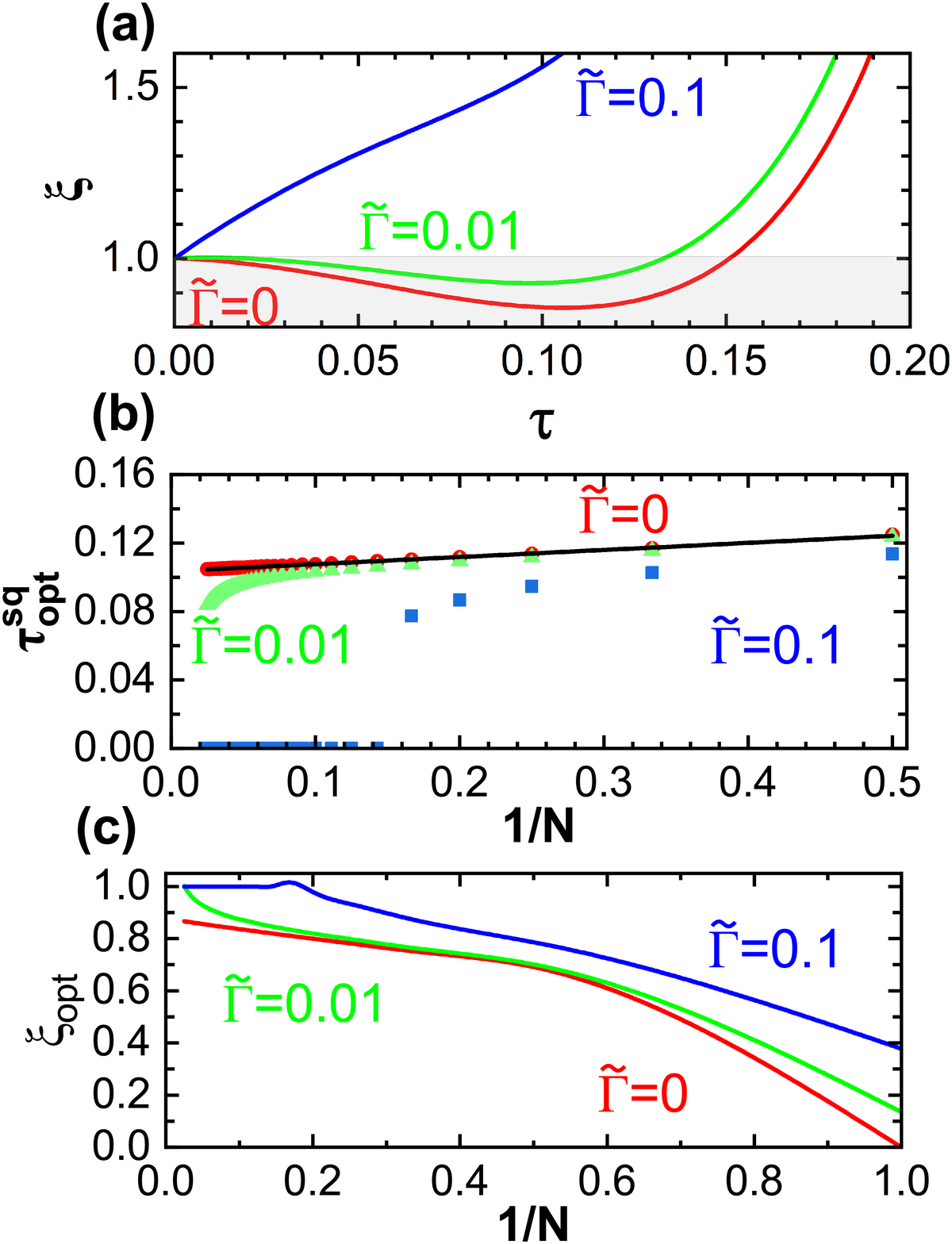}
\caption{The Wineland squeezing parameter for the state (\ref{rhosol}) with decoherence. (a) Wineland squeezing parameter versus QND interaction time $ \tau $ with decoherence rates $ \tilde{\Gamma} $ as marked.  The atom number is $ N = 20 $.  (b) The optimal squeezing time $\tau_{\text{opt}} $ determined by minimizing $\xi(\tau)$ plotted as a function of $ 1/N $. A fit of the optimal times for $\Gamma=0$ is given by $\tau_{\text{opt}}^{(\tilde{\Gamma}=0)}=0.104+\frac{0.0413}{N} $.  (c) The optimal Wineland squeezing $\xi (\tau_{\text{opt}}) $ versus $1/N$ for different decoherence rates $ \tilde{\Gamma} $. 
%[TB: I don't think we need this]
%\textcolor{red}{The numerical results of optimal Wineland squeezing for $\tilde{\Gamma}=0$  show a good coincidence with Pad{e'} approximation ${\xi}_{\text{opt}}^{\tilde{\Gamma}=0 }  =\frac{0.877\, -\frac{0.877}{N}  }{-\frac{0.392}{N^2}  -\frac{0.537}{N} +1}$.} 
For all plots, we choose parameters $n_c=100,n_d=0,\alpha=10$.  }
\label{wineland plot}
\end{figure}

Figure \ref{wineland plot}(a) shows the Wineland squeezing parameter (\ref{wineland squeezing}) as a function of time $\tau$ for different decoherence factors $\tilde{\Gamma} $. For small dephasing rates, the Wineland squeezing parameter detects entanglement (shaded region) in the short time regime.  In comparison to the Hofmann-Takeuchi and EPR steering criterion as calculated in Ref. \cite{Gao_2022}, the region is reduced, indicating that this is a less sensitive detector of entanglement. As was found in Ref. \cite{Gao_2022}, entanglement is present for all times $ 0 < \tau < \pi $ for dephasing rates $ \Gamma = 0, 0.01, 0.1 $ as detected by logarithmic negativity.  The various entanglement witnesses are only a sufficient condition for entanglement, and a lack of violation of the criterion does not indicate that entanglement is not present.  For small dephasing rates, there is an optimal interaction time $ \tau_{\text{opt}}$ which minimizes the Wineland squeezing parameter (Fig. \ref{wineland plot}(b)).  This shows generally a linear relationship with $ 1/N $.  In Fig. \ref{wineland plot}(c), the relationship between optimal Wineland squeezing parameter $\xi_{opt} = \xi ( \tau_{\text{opt}}) $ with atom numbers $1/N$ are shown for various dephasing rates. 
We observe that $\xi_{opt}$ remains in the entanglement detection region for the dephasing rates $ \tilde{\Gamma} = 0,0.01 $. For $ \tilde{\Gamma} = 0.1 $, $ \xi_{\text{opt}}  =1 $ for larger values of $ N $ and hence no entanglement is detected.  For this regime, the dephasing is strong enough to remove the minimum as seen in Fig. \ref{wineland plot}(a).  Hence we see that it becomes more difficult to detect entanglement using the Wineland squeezing criterion for large ensembles and larger dephasing rates.  Hence as an entanglement detection tool, other methods such as the Hoffman-Takeuchi criterion appear to be a better choice.

\subsubsection{Bell-CHSH inequality}
\label{szdephasingchsh}

\begin{figure}[t]
\centering
\includegraphics[width=\columnwidth]{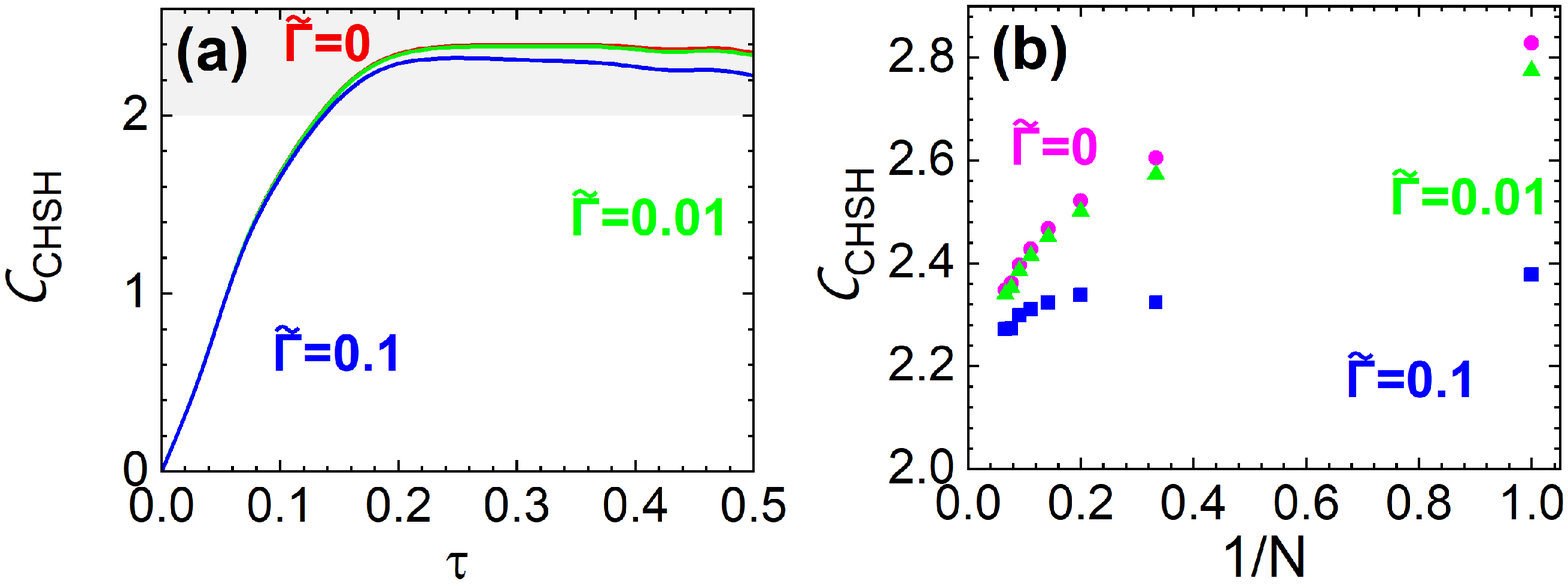}
\caption{Bell-CHSH correlations for the state (\ref{rhosol}) with atomic dephasing. (a) Time dependence of (\ref{bell-chsh}) for different decoherence ratios as marked. Here we choose parameters $\theta_{B}=0.251,N=11$. (b) Optimal values of (\ref{bell-chsh})  with respect to $\tau$ and $\theta_{B}$ for various $N$. Common parameters throughout are $n_{c}=100,n_{d}=0,\alpha=10$.  }
\label{chshdephasing}
\end{figure}

In Fig. \ref{chshdephasing}(a), we calculate the left-hand side of the Bell-CHSH inequality (\ref{bell-chsh}) for the state (\ref{rhosol}) as a function of the QND interaction time with different dephasing rates.
For each dephasing rate, there is a slightly different dependence with $ \tau $, where larger dephasing rates tend to diminish the violation for longer times.  
Fig. \ref{chshdephasing}(b) shows the  optimized value of (\ref{bell-chsh}) as a function of $N$. We see that the Bell violations are observed for all $N$ even in the presence of dephasing. The surprisingly robust nature of the Bell violations in the presence of dephasing is a positive sign that such correlations may be observable in experimental situations.  We note however that the violations tend to diminish for larger ensembles so that for larger systems, such that alternative criteria such as the Hoffman-Takeuchi criterion may be more sensitive detectors of entanglement.

\section{Conclusion}
\label{conclusion}

We have studied the effect of three types of decoherence (optical phase diffusion, photon loss and gain, and atomic dephasing) on entangled states produced by QND measurements in atomic ensembles.  Generally, we find a similar conclusion to Ref. \cite{Gao_2022} where the states are relatively robust for all types of decoherence in the short time regime.  For example, in the entanglement plots of Fig. \ref{entanglment plots}, even for very large decoherence rates there is a significant amount of entanglement in the short time regime but decays for longer times.  The correlations, as shown in Figs. \ref{density1},\ref{kuosanfangcha},\ref{laserfangcha} are all robust in the short time regime but can degrade at longer interaction and decoherence times.  As such, correlation-based detection of entanglement and EPR steering tend to be robustly observed in the short time regime as seen in Figs. \ref{kuosanentanglment criteria},\ref{laserfangcha},\ref{wineland plot}.  Interestingly, Bell correlations are also very robust under various types of decoherence, as seen in Figs. \ref{allchsh},\ref{chshdephasing}. The robustness of Bell's violation can be attributed to the fact that the optimal interaction time occurs for the short time regime.  The main issue here with experimental observation is that for larger $ N $ the level of violation diminishes, making its detection more challenging.   

Our results suggest that producing entangled states based on QND measurements should be one of the most experimentally viable methods in the context of BECs.  As mentioned in the introduction, while such methods are well established for entanglement generation in hot atomic ensembles \cite{julsgaard2001experimental,hammerer2010quantum}, there is no corresponding experiment that has been realized for BECs.  While there have been numerous approaches that have been theoretically investigated for entanglement generation in BECs already \cite{treutlein2006microwave,bec7,bec6,2013Entanglement,2014Entanglement,bec5,hussain2014,abdelrahman2014coherent}, so far none have been realized experimentally.  The QND approach allows for a robust and versatile method to generate such entanglement.  The theory that has been developed in \cite{AristizabalZuluaga2021QuantumNM} and summarized in this paper allows for the investigation beyond the short time regime, to produce more exotic types of entanglement.  The generation of such highly entangled many-body states is of fundamental interest to the macroscopic nature of such states, and also has been shown to have applications in quantum information \cite{pyrkov2014quantum,pyrkov2014full,ilo2018remote,PhysRevA.103.062417,byrnes2012macroscopic,BYRNES2015102}.

%We observe that the effect of photon loss on the quantum state is to partially mask the present entanglement between BECs. This is brought about by the photon loss causing  $\text{Var}(S^x_1 - S^x_2)$ and $\text{Var}(S^y_1 + S^y_2)$ to approach the standard quantum limit normalized to unity. Similar effects are found in probability distribution which is initially correlated in the $S^x$ basis with a zero width. In the presence of photon loss, correlation is redistributed such that the probability density distribution in $S^x$ acquires a finite width. While photon loss blurs the entanglement in the quantum state, the amount of entanglement in the quantum state is unchanged from its value without photon loss. This is verified by the logarithmic negativity which has the same value irrespective of the photon loss present in the system. 

\section*{Acknowledgments}

This work is supported by the National Natural Science Foundation of China (62071301); NYU-ECNU Institute of Physics at NYU Shanghai; the Joint Physics Research Institute Challenge Grant; the Science and Technology Commission of Shanghai Municipality (19XD1423000,22ZR1444600); the NYU Shanghai Boost Fund; the China Foreign Experts Program (G2021013002L); the NYU Shanghai Major-Grants Seed Fund; Tamkeen under the NYU Abu Dhabi Research Institute grant CG008.

\section*{Data availability statement}

The data that supports the findings of this study are available within the article (and any supplementary material).

\appendix

\section{Density matrix for optical phase diffusion}
\label{app:phasediff}

In this section, we give details of the explicit expression for the density matrix including optical phase diffusion.  Starting from the initial state (\ref{initial density matrix}), we evaluate 
\begin{align}
& \rho_{\text{PD}} = \langle n_c | \langle n_d | \sum_{m,n,m',n'=0}^{\infty} M_{m, n}^{\text{PD},c}  (t)  M_{m', n'}^{\text{PD},d}  (t) \nonumber \\
& \times \rho_{0} {M_{m', n'}^{\text{PD},d}}^{\dagger} (t) 
 {M_{m, n}^{\text{PD},c}}^{\dagger} (t) | n_c \rangle | n_d \rangle   \nonumber \\ 
& =\frac{1}{4^{N}}\sum_{k_1,k_2,k_1',k_2'=0}^{N}  \sqrt{\binom{N}{k_1}\binom{N}{k_2}
\binom{N}{k_1'} \binom{N}{k_2'}} z^{2}(\tau)\nonumber\\
& \times e^{- |\alpha |^2}e^{u(\tau){ |\alpha|}^{2}\cos[(k_{1}-k_{2}-k_1'+k_2')\tau] } 
\nonumber \\
 &\times z^{2(n_c-m+n_d-m')}(\tau)  \sum_{m = 0}^{n_c} \sum_{m' = 0}^{n_d}\frac{u^{m+m'}(\tau)}{m!m'!}  \nonumber \\
& \times \left \{\alpha^2
\cos[(k_1-k_2)\tau] \cos[(k_1'-k_2')\tau ] \right \}^{n_c-m} \nonumber \\
&\times \left \{\alpha^2
\sin[(k_1-k_2)\tau] \sin[(k_1'-k_2')\tau ] \right \}^{n_d-m'} \nonumber \\ 
&\times \frac{n_{c}!n_{d}!}{[(n_{c}-m)!]^{2}[(n_{d}-m')!]^{2}} 
 | k_1\rangle | k_2\rangle \langle k_1'| \langle k_2'| \nonumber \\
 &= \frac{1}{4^{N}}\sum_{k_1,k_2,k_1',k_2'=0}^{N}  \sqrt{\binom{N}{k_1}\binom{N}{k_2}
\binom{N}{k_1'} \binom{N}{k_2'}}  \nonumber \\
&\times \frac{[-u(\tau )]^{n_c} U\left(-{n_c},1,-\frac{{n_c}
 \cos [({k_1}- {k_2}) \tau ] 
\cos [({k_{1}'}-{k_{2}'}) \tau ] z^2(\tau )}{u(\tau )}\right)}{n_c!}\nonumber\\
&\times \frac{[-u(\tau )]^{n_d} U\left(-{n_d},1,-\frac{{n_d}
 \sin [({k_1}- {k_2}) \tau ] 
\sin [({k_{1}'}-{k_{2}'}) \tau ] z^2(\tau )}{u(\tau )}\right)}{n_d!}\nonumber\\
&\times z^{2}(\tau)
D[(k_1-k_2-k_1'+k_2')\tau]
| k_1\rangle | k_2\rangle \langle k_1'| \langle k_2'| . 
\label{phasediffdensmat}
\end{align}
Here 
\begin{align}
   z\left ( \tau \right ) & = \frac{1}{\tilde{\kappa} \tau+1} \nonumber \\
   u(\tau) & =\frac{\tilde{\kappa} \tau}{\tilde{\kappa} \tau +1} \nonumber \\
   D(\chi )& =e^{-|\alpha|^2[1-u(\tau)\cos(\chi)]}
\end{align}
and $U$ is the confluent hyper-geometric function of the second kind.

\section{Fock state matrix elements}
\label{app:fockxyz}

The matrix elements of the $S^y$ rotation is given by \cite{byrnes2020quantum}
\begin{align}
& \langle k | e^{-i S^y \theta/2} | k' \rangle = \sqrt{ k'! (N-k')! k! (N-k)!} \nonumber \\
& \times 
\sum_n \frac{(-1)^n \cos^{ k- k' + N - 2n} (\theta/2) \sin^{2n + k' - k} (\theta/2) }{(k-n)!(N-k'-n)!n!(k'-k+n)!}, 
\label{syrotmatrixelement}
\end{align}
where $ | k \rangle $ are the eigenstates of $ S^z $. The matrix elements of $ S^x $ are accordingly given by
\begin{align}
& \langle k | e^{-i S^x \theta/2} | k' \rangle = 
i^{k'-k} \langle k | e^{-i S^y \theta/2} | k' \rangle,
\label{sxrotmatrixelement}
\end{align}
using the fact $ S^x = e^{-i S^z \pi/4} S^y e^{i S^z \pi/4} $.

\section{Density matrix for photon loss and gain}
\label{app:lossgain}

In this section, we give the explicit expression of  the density matrix including photonic gain and loss.  Starting from the initial state (\ref{initial density matrix}), we evaluate 
\begin{align}
& \rho_{\text{LG}} = \langle n_c | \langle n_d | \sum_{p,q,p',q'=0}^{\infty} M_{p, q}^{\text{LG}}  (c)  M_{p', q'}^{\text{LG}}  (d) \nonumber \\
& \times \rho_{0} {M_{p', q'}^{\text{LG}}}^{\dagger} (d) 
 {M_{p, q}^{\text{LG}}}^{\dagger} (c) | n_c \rangle | n_d \rangle  \nonumber \\
&=\frac{1}{4^N} \sum_{k_1,k_2,k_1',k_2'=0}^{N} \sqrt{\binom{N}{k_1}
 \binom{N}{k_2}\binom{N}{k_1'}\binom{N}{k_2'}} \nonumber \\
&{T_3}^{2}R[(k_1-k_2-k_1'+k_2')\tau ] 
\sum_{p=0}^{n_{c}}\sum_{q=0}^{n_{d}} \frac{(g T_1)^{p+q}}{p!q!}   \nonumber \\
&\times \{\alpha^2 \cos[(k_1-k_2)\tau]\cos[(k_1'-k_2')\tau]\}^{n_c-p} \nonumber \\
&\times \{\alpha^2 \sin[(k_1-k_2)\tau]\sin[(k_1'-k_2')\tau]\}^{n_d-q} \nonumber  \\
&\times {T_2}^{2(n_c-p+n_d-q)} \frac{n_c!}{{[(n_c-p)!]}^2} \frac{n_d!}{{[(n_d-q)!]}^2}
| k_1\rangle | k_2\rangle \langle k_1'| \langle k_2'|\nonumber \\
&= \frac{1}{4^N} \sum_{k_1,k_2,k_1',k_2'=0}^{N} \sqrt{\binom{N}{k_1}
 \binom{N}{k_2}\binom{N}{k_1'}\binom{N}{k_2'}}\nonumber\\
&\frac{(-g{T_1})
^{{n_c}}U\left(-{n_c},1,-\frac{
\alpha ^2\cos  [({k_1}-{k_2})\tau]
\cos  [({k_1'}-{k_2'})\tau]
{T_2}^2}{g{T_1}}\right)}{{n_c}!}\nonumber\\
&\frac{(-g{T_1})
^{{n_d}}U\left(-{n_d},1,-\frac{
\alpha ^2\sin  [({k_1}-{k_2})\tau]
\sin  [({k_1'}-{k_2'})\tau]
{T_2}^2}{g{T_1}}\right)}{{n_d}!}\nonumber\\
&\times{T_3}^{2}R[(k_1-k_2-k_1'+k_2')\tau ]| k_1\rangle | k_2\rangle \langle k_1'| \langle k_2'| , 
\label{laser density}
\end{align}
where $R(\chi)=e^{-|\alpha|^2[1-\gamma T_1\cos(\chi)]}$.

\bibliographystyle{apsrev}
\bibliography{gaoshuai}
%% 2) Copy the. bbl file to below and comment out the above two lines. 

\end{document}